\documentstyle[12pt]{article}
\begin{document}
\font\ninerm = cmr9

\def\footnoterule{\kern-3pt \hrule width \hsize \kern2.5pt}

\pagestyle{empty}
\hfill gr-qc/9603014
\begin{center}
{\large\bf Limits on the Measurability of Space-time Distances in}

{\large\bf (the Semi-classical Approximation of) Quantum Gravity}%
\footnote{\ninerm This work is supported
in part by funds provided by the U.S. Department of Energy (D.O.E.)
under cooperative agreement \#DE-FC02-94ER40818,
and by Istituto Nazionale di
Fisica Nucleare (INFN, Frascati, Italy). }

\vskip 1cm
Giovanni AMELINO-CAMELIA
\vskip 0.5cm
{\it Center for Theoretical Physics\\
Laboratory for  Nuclear Science and Department of Physics\\
Massachusetts Institute of Technology\\
Cambridge, Massachusetts 02139, U.S.A.}

\end{center}

\vspace{1.2cm}
\begin{center}
{\bf ABSTRACT}
\end{center}

{\leftskip=0.6in \rightskip=0.6in

By taking into account 
both quantum mechanical and general relativistic
effects,
I derive an equation that
describes some
limitations on the measurability of space-time distances.
I then discuss possible features of quantum gravity which
are suggested by this equation.

}

\vskip 2cm
\centerline{{\it Modern Physics Letters} A9 (1994) 3415}

\vfill

\hbox to \hsize{INFN-NA-94/35 \hfil September 1994}

\newpage

\baselineskip 12pt plus .5pt minus .5pt
\pagenumbering{arabic}
\pagestyle{plain} 

\section{Introduction}
The study of problems in which both quantum mechanical (QMal) and general
relativistic (GRic) effects are important is strongly motivated
by the possibility of discovering some
features of Quantum Gravity (QG),
a yet to be found ``more fundamental" theory
in which the questions raised by the apparent incompatibilities of 
Quantum Mechanics (QM) and General Relativity (GR)
are solved.
One such problem that has recently attracted a lot of attention is 
the (semi-classical) quantum analysis of black holes, which has 
lead to the famous {\it black hole information paradox}\cite{bh}.
In this Letter I shall be concerned with 
another extensively investigated 
problem that, in most treatments, 
involves the interplay between QMal and GRic effects: the search for 
limitations on the quantum measurement  of space-time distances
(see, 
for example, Refs.[2-8]; in particular, Ref.\cite{engl} 
gives a good review of the results in this area
and refers to additional related
material).
Specifically, I shall analyze from a different point of view
the interesting measurement procedure for space-time distances
discussed in Refs.\cite{wign,ng}, and derive a simple equation that
indicates some possible features of QG.

\section{Measurement Procedure}
In ordinary QM the
distance between two space-time points is 
given in terms of
their coordinates, which are assumed to have a physical meaning
independent of observations
(no prescription for the measurement of these coordinates is to be given).
In GR, instead, space-time points can only be meaningfully identified
by events.
Therefore, a basic measurement to be considered
in (at least the semiclassical limit of)\footnote{Here
and in the title I point out that my discussion might only apply 
to the semiclassical
limit (GRic geometry and quantized matter) of QG.
In fact, it is possible that, at scales smaller than the Plank length, 
the very concept of distance may be somewhat foreign 
to the full QG because geometry
might look very different from what we are accustomed to\cite{engl}.}
QG is the quantum measurement of the distance between events in space-time.
In particular, it is important to establish whether there
are limitations on the accuracy achievable in these
measurements, because such limitations would render impossible
the definition of a traditional coordinate system\cite{wign}.
A simple way to start analyzing this issue 
is given by the study of the measurement of the length $L$ of an 
object, 
which was considered in Ref.\cite{ng}.
Clearly, this 
does not lead to the most general discussion of the
quantum measurement of the distance between events in space-time
(in particular, the very idea of
the ``length of an   
object" can only be introduced in approximately flat regions of space-time);
however, its simplicity allows a very intuitive analysis, which is
useful in deriving expectations for the features of QG.
As discussed in Refs.\cite{wign,ng}, in the spirit of GR
such a length measurement
can be carried out by putting a clock, a ``light-gun" ({\it i.e.} a device 
capable of sending
a light signal when triggered), and a detector
at one end of the object and putting a mirror at the other 
end. [In the discussions of Refs.\cite{wign,ng} the light-gun
and the detector are not explicitly mentioned; however, their ``clock"
is capable of sending the signal at $t \! = \! 0$ ({\it i.e.} it is connected
to a light-gun), and stops when hit by the signal coming back from the mirror
({\it i.e.} it can detect the signal).]
This apparatus should be set to send a light signal toward the mirror
when the clock reads zero, and to record
the time $T$ shown by the clock when
the light signal is detected by the detector after being
reflected by the mirror.
Clearly the time $T$ is related to
the length $L$; for example,
in Minkovski space and neglecting quantum effects one finds that
$T$ and $L$ are simply related by
\begin{eqnarray}
L = c {T \over 2} ~,
\label{naivel}
\end{eqnarray}
where, as customary, $c$ denotes the speed of light.

I am interested in analyzing this measurement
procedure including QMal and GRic effects, and therefore
the relation between $T$ 
and $L$ is more complex than Eq.(\ref{naivel}).
Specifically, the outcome 
$T$ (time read by the clock) of a measurement procedure
and $L$ (the ``actual" length of the object) are formally
related as follows
\begin{eqnarray}
L = c {T \over 2} \pm \delta L + \Delta L \pm \delta_g L ~,
\label{general}
\end{eqnarray}
where 
\begin{itemize}
\item{} $\delta L$ is the total QMal {\bf uncertainty} due to 
the QMal uncertainties in the position and velocity of the various
agents in the measurement procedure.
As it was already discussed in
Refs.\cite{wign,ng},
contributions to $\delta L$ come, for example, from the spread 
in the position of the various devices
(clock, light-gun, detector, and
mirror) during the time interval $T$.
\item{} $\Delta L$ is the total classical
{\bf correction} (to the relation $L = c T / 2$) due to the 
gravitational forces among the 
agents in the measurement procedure.
Examples of such corrections are the ones resulting from the
gravitational attraction between the light signal 
and the devices in the apparatus.
\item{} $\delta_g L$ is the total QMal {\bf uncertainty} 
which results from the uncertainties in the
gravitational forces among the 
agents in the measurement procedure.
For example, as a result of the QMal spread in the mass of the clock, 
there is an uncertainty in the strength of the
gravitational attraction exerted by the clock on the light signal.
\end{itemize}

In this Letter I 
intend to investigate, as intuitively as possible,
the possibility of 
limitations on the accuracy of quantum measurements of $L$
resulting from the fact that 
it is not possible to prepare the apparatus
so that
the uncertainty introduced in the measurement
of $L$ by the presence of the $\delta$'s in Eq.(\ref{general}) be 
arbitrarily small.

\noindent
I shall not attempt to derive {\bf the absolute lower bound}
(whose rigorous derivation 
surely involves an extremely complex analysis) on the
uncertainty of quantum measurements of $L$, 
instead I shall look for {\bf a lower bound} (which may well be lower than 
the absolute lower bound) for this uncertainty.

\noindent
Consistently with this objective, 
the only contribution to $\delta L$ 
that I will consider is $\delta x_{com}$, defined as the spread 
in the position of the center of mass of the system composed
by clock, light-gun, and detector
during the time interval $T$. 
The other (many!) contributions to $\delta L$ (given, among others,
by the spread in the position of the
clock, light-gun, and detector with respect to the center of mass
of the clock+light-gun+detector system,
and by the spread in the position of the
mirror) could obviously only increase 
the lower bound
on the uncertainty that I will present, but, like the authors 
of Ref.\cite{ng}, I prefer to give an intuitive and simple
discussion rather than attempting to find a higher lower bound.

\noindent
$\delta_g L$
has already been considered in Refs.\cite{padma,engl}, 
where it has been found that 
$\delta_g L \geq L_p$, $L_p$ being the Plank length.
Therefore, in the following, I shall look for a lower 
bound $\delta x^{(min)}_{com}$
for $\delta x_{com}$, which will lead to a lower bound for the
uncertainty $\delta L^{tot}$
in our length measurement, as indicated by the relation
\begin{eqnarray}
\delta L^{tot} \equiv \delta L + \delta_g L \geq
\delta x_{com} + L_p \geq \delta x^{(min)}_{com} + L_p
~.
\label{defdltot}
\end{eqnarray}

\noindent
In the following, I shall also assume 
(these simplifying assumptions will allow me to use spherical symmetry)
that the
clock+light-gun+detector system 
has (homogeneously distributed) mass $M$ 
and occupies  
a spherically symmetric region of space of radius
(``size") $s$,
and neglect the effects of all the other masses in the problem\footnote{As
discussed in Ref.\cite{ng}, accounting for the effects of the other
masses (and accounting for the 
the QMal spread in the mass $M$)
could only increase the lower limit on $\delta L^{tot}$ found in this
type of analysis.}.

Finally let me observe that 
the $\Delta L$ contribution to Eq.(\ref{general})  does not play 
any role in my study because, as I indicated, 
in a context in which both GRic and QMal effects are taken into account,
$\Delta L$ represents a correction, not an uncertainty. In fact, knowing all 
the masses in the system, $\Delta L$ can be calculated and accounted for
in the analysis of the outcome of the measurement, therefore leading
to no additional uncertainty.

\noindent
In Ref.\cite{ng} $\Delta L$ was essentially treated on the same footing
as the $\delta$'s. 
This makes the results of
Ref.\cite{ng} relevant to contexts different from the one that I am 
here considering;
they apply to measurements in which  
the masses in the apparatus are not known, or
to measurements whose outcome is analyzed ignoring
GRic effects (in which case $\Delta L$ is to be treated as an experimental
error). 

\section{Analysis}

\subsection{Quantum Mechanics}

The evaluation of the spread 
in the position of (the center of mass of) a body 
during a time interval $T$ is an elementary QMal problem.
Following Ref.\cite{wign}, one finds that 
\begin{eqnarray}
\delta x_{com} \equiv \delta x_{com}(t \epsilon [0,T]) 
\geq \delta x_{com,i} 
+ { \hbar \over c } 
{ 2 L \over M \delta x_{com,i} } 
~,
\label{dawign}
\end{eqnarray}
where $\delta x_{com,i} \equiv \delta x_{com}(t=0)$ is 
the initial ({\it i.e.} at the time $t\!=\!0$ when the 
light signal is emitted)
spread in the position 
of the center of mass of
the clock+light-gun+detector system.

Eq.(\ref{dawign}) can be understood as follows\cite{wign,ng}.
Initially, the system is described by a wave packet with
position-spread $\delta x_{com,i}$ and velocity-spread
$\delta v_{com,i}$. 
During the time interval $T$
following $t\!=\!0$ the uncertainty in the position 
of the center of mass of
the clock+light-gun+detector system is given 
by
\begin{eqnarray}
\delta x_{com} \sim \delta x_{com,i} + \delta v_{com,i} \, T 
\sim \delta x_{com,i} + \delta v_{com,i} \, { 2 L \over c} 
~,
\label{dawign2}
\end{eqnarray}
where on the right-hand-side I used the fact that in first
approximation $T \sim 2 L/c$.

Eq.(\ref{dawign2}) takes the form of Eq.(\ref{dawign}) 
once the uncertainty principle, which states that
\begin{eqnarray}
\delta x_{com,i} \, \delta v_{com,i} \geq {\hbar \over M}
~,
\label{up}
\end{eqnarray}
is taken into account.

The most important feature of Eq.(\ref{dawign}) is that it indicates that
for a clock+light-gun+detector
system of a given mass $M$ there is no way to prepare the $t=0$-wave-packet
so that $\delta x_{com} = 0$. In fact, 
Eq.(\ref{dawign}) indicates that, for given $M$, QM leads to the following
minimum value of $\delta x_{com}$ 
\begin{eqnarray}
\left[ \delta x_{com}^{(min)}\right]_{QM} \sim 
\sqrt{{\hbar \over c} {L\over M}}
~,
\label{dminqm}
\end{eqnarray}
where, as I shall do in the following, 
I neglected numerical factors of $O(1)$, which are essentially
irrelevant for the discussion presented in this Letter.

\subsection{General Relativity}

Up to this point I have only used QM 
and therefore it is not surprising to
discover that, as shown by Eq.(\ref{dminqm}), 
$[\delta x_{com}^{(min)}]_{QM} \! \rightarrow \! 0$ as
$M \! \rightarrow \! \infty$. Indeed, 
the uncertainty that I am considering originates from
the uncertainty in the kinematics of the bodies involved
in the measurement procedure,
and clearly this uncertainty vanishes in the limit
of infinite masses\footnote{This observation plays for example
a crucial role in Bhor and Rosenfeld's\cite{rose} measurement analysis
concerning quantum electrodynamics. Specifically, it motivates the choice,
as an agent in their gedanken experiment,
of a continuous charge distribution whose ratio of electric charge versus mass
can be taken to zero.
Limitations, like the ones that I discuss in the following, 
on the measurability of spacetime distances in QG
are just due to the fact that the QG-charge and the mass
are the same thing (equivalence of inertial and gravitational mass), 
their ratio is fixed to 1, and therefore the type of measurement
strategy adopted by Bhor and Rosenfeld is not viable in QG.}
(the classical limit) because in this limit Eq.(\ref{up}) is
consistent with $\delta x_{com,i} \! = \! \delta v_{com,i} \! = \! 0$.

\noindent
However, the central observation of the present paper is that
this scenario is significantly modified, 
as a result
of the dramatic consequence on the geometry of large ``localized" (what
I mean here with localized will become clear later) masses,
when the GRic effects relevant to our experimental set up are 
taken into account.

For the moment, let us consider
a fixed size $s$ for the
clock+light-gun+detector system. 
Large values of the mass of the
clock+light-gun+detector system necessarily lead to great
distorsions of the geometry, and well before the $M \! \rightarrow \! \infty$
limit (which, as I just indicated, is desirable
for reducing the uncertainty given by Eq.(\ref{dminqm})) our
measurement procedure can no longer be followed.
In particular, if 
\begin{eqnarray}
M \geq c^2 {s \over G} = {\hbar \over c} {s \over L_p^2}
~,
\label{mhor}
\end{eqnarray}
where $G$ is Newton's constant of gravitation, an {\it horizon} forms around 
the center of mass of the
clock+light-gun+detector system
(here I am using the spherical symmetry) and it is not possible
to send a light signal from the clock+light-gun+detector system
to the mirror positioned
on the other side of the object whose length is being 
measured\footnote{If (at a given moment)
there is a device inside the horizon of a black hole 
and close to the center, and another device further 
away from the center (it does not matter whether this second device
is inside or outside the horizon as long as it is at a greater distance 
from the center than the other device),
signals emitted by the device
which is closer to the center (the clock+light-gun+detector system)
cannot reach the device which is further away (the mirror).

\noindent
I thank Mario Bergeron and John Stachel for discussing with me about this.}.
This observation, combined with Eq.(\ref{dminqm}),
allows me to derive the following relation giving
a $L$-dependent lower bound on the uncertainty 
\begin{eqnarray}
\delta L^{tot} \geq
\delta x_{com}^{(min)} + L_p \sim \sqrt{L L_p^2 \over s} + L_p
~.
\label{dmin}
\end{eqnarray}
It must be noted that $\delta x_{com}^{(min)} \rightarrow 0$ in 
the $s \rightarrow \infty$ limit;
however, 
it is rather unclear whether it makes any sense
to allow that in our experimental set-up 
the clock+light-gun+detector system 
has infinite size (or even that  $s \! > \! L$)
\footnote{Moreover, it is easy to realize\cite{amelprep} that
(even if we allow the size of the clock+light-gun+detector system
to be extremely large) the lower bound on $\delta L^{tot}$
might still be rather strongly $L$-dependent because the size $s$ which
is in fact relevant in the determination of a significant lower bound
is the size of the region of the 
clock+light-gun+detector system which 
is actually interacting with the photon.}.

\section{Conclusion}
\renewcommand{\theequation}{\Roman{section}.\arabic{equation}}

In this last section, I want to discuss some
possible features of QG which might be indicated by Eq.(\ref{dmin}).

\subsection{Classical Device}

My first observation is that
Eqs.(\ref{dminqm}) and (\ref{dmin}) can be interpreted\cite{amelprep} as
indications of the fact that whereas 
in QM a {\it classical device}\footnote{With {\it classical device}
I intend a device
(an example of which would be given in my measurement analysis by the 
clock+light-gun+detector system if it was infinitely massive and extended)
which can perform observations ({\it i.e.} can probe with signals the
system under measurement), and
whose position and velocity are both completely
determined, so that the accuracy of its observations is only
limited by the application of the uncertainty principle to the quantities
observed. In particular, such a device should 
be able to measure distances (or lengths) with unlimited accuracy (obviously
at the price of renouncing to any information concerning momenta), at least
up to scales of order $L_p$.}
is ``only" required to be
infinitely massive, 
in (at least the semiclassical approximation of) QG
a {\it classical device} is required to be
infinitely massive and infinitely extended.

\noindent
It would be interesting to investigate whether this observation is compatible
with our present measurement theory, in which the presence 
of {\it classical devices} is a crucial ingredient.

\subsection{Decoherence}

An interesting possibility suggested by Eq.(\ref{dmin})
is the one of decoherence. 
In fact, Eq.(\ref{dmin}) indicates that 
in every {\it real-world} ($s \! \ne \! \infty$) experiment,
unless $s\! \sim \!L$,
the uncertainty $\delta L^{tot}$ 
on the measurement of $L$ depends on $L$ itself,
and, as discussed for example in Refs.\cite{karo,ng}, this should lead
to quantum
(de)coherence phenomena.

\noindent
It would not come as a surprise if indeed QG would host a
mechanism for decoherence; in particular, this has already been speculated
in Refs.\cite{karo,ng}
and in some approaches to the {\it black hole information paradox}.
Moreover, decoherence has also been advocated in discussions of
the QMal ``measurement problem", and
if QG (besides easying the tension between QM and GR) must also give us
a satisfactory measurement theory, one might expect 
decoherence to be present\footnote{On  this issue of the relation between
Eq.(\ref{dmin}), decoherence, and the QMal ``measurement problem"
let me make parenthetically a speculative but perhaps intriguing
comment.
In the QG measurement analysis presented in Ref.\cite{bergstac}
it has been argued that in any such measurement
procedure the information 
must be first transferred from the system being measured to some intermediate
microscopic system (also to be considered part of the apparatus)
before being recorded by the appropriate macroscopic devices.
In such a scenario one can see\cite{amelprep} 
that the size $s$ relevant to 
the lower bound on the uncertainty of length measurements 
that I have discussed is the size
of such microscopic systems, and therefore the ``amount of decoherence"
would be rather large.

\noindent
If QG is to give us a consistent measurement theory, one can also envision
that QG might prescribe a typical size $s^{*}$
(which could be related to or 
be given by 
the Plank length $L_p$, which is the natural length scale in the problem)
for the kind
of microscopic systems which can appear in the very first
stage of the measurement procedure. In this scenario, once substituted
$s^{*}$ to $s$, Eq.(\ref{dmin})
would give an absolute and general lower bound on the accuracy of
measurements of space-time distances in QG.}.

\subsection{Material Reference Systems}

In Ref.\cite{rovelli}, Rovelli has identified well-defined local
observables of QG, by explicitly taking into account the physical
nature of the bodies that form the reference system,
which is understood as a {\it material reference system}.
A rather typical {\it material reference system} is given by a collection of
bodies of size $\tilde{s}$, and one of the local observables available in such
a {\it material reference system} is the distance between two of the bodies.

\noindent
Eq.(\ref{dmin}) appears to be immediately relevant to this framework.
In particular, it indicates that the distance between two of the bodies
of a {\it material reference system} cannot  be measured with infinite
accuracy, the uncertainty being bounded from below 
by $L_p \! + \! \sqrt{L L_p^2 / \tilde{s}}$; 
moreover, decoherence phenomena should characterize the physics
as described by one such {\it material reference system}.

Also notice that in this case the $\tilde{s} \rightarrow \infty$ limit
is meaningless\footnote{It is not possible to
set up the network of bodies necessary to form
the {\it material reference system}
if each one of the bodies occupies all of space.
Moreover, Rovelli's observables are defined only with respect to a given
{\it material reference system}, {\it i.e.} they are characterized by 
a specific value of $\tilde{s}$, and therefore increasing $\tilde{s}$
one would not reach better and better accuracy in the measurement of
{\bf the same} Rovelli's observable, 
one would instead span a class of {\bf different} Rovelli's
observables which are measurable with different accuracy depending on their
specific value of $\tilde{s}$.}, 
and
actually, in order to have as fine as possible a
network of bodies, one would like $\tilde{s}$ to be small, but this,
following Eq.(\ref{dmin}),
leads to large uncertainty in length measurements and large decoherence.

\subsection{Outlook}

In conclusion, ``probing" (even if only conceptually)
an area of physics where both QMal and GRic effects can be relevant
has once again lead to an intruiging result.
It appears that
the analysis of procedures of measurement of space-time distances
holds promises of being instructive not only on the contradictions
between QM and GR, but also on the unsolved issues of the QMal measurement
theory, and therefore
there are reasons of interest in pursuing further
the type of investigation
preliminarily presented,
in their respective limits of
validity,
in this Letter and in Ref.\cite{ng}.
In particular, it would be important to verify whether the results
are relevantly modified by removing the simplifying assumptions made
here and in Ref.\cite{ng}, which allow a spherically symmetric treatment.
It would also be interesting to estimate whether the decoherence phenomena
to be expected, as a result of Eq.(\ref{dmin}), 
in {\it real-world} laboratories 
(with their typical sizes) 
are large enough to be observed.  

\bigskip
\bigskip
\bigskip
\bigskip
In the preliminary stages of this work I greatly benefitted from
very stimulating discussions with John Stachel, 
which I am very happy to acknowledge.
I am also happy to acknowledge conversations with 
Jack Ng, with whom I discussed issues related to Ref.\cite{ng}, and
several members
of the Center for Theoretical Physics, especially Dongsu Bak,
Mario Bergeron,  
Domenico Seminara, and Philippe Zaugg.

\newpage
\baselineskip 12pt plus .5pt minus .5pt

\end{document}